\documentclass[aps,prl,twocolumn,amsmath,amssymb,bibnotes,longbibliography,10pt]{revtex4-1}
\usepackage[utf8]{inputenc}
\usepackage{graphicx}
\usepackage{textcomp}
\usepackage{pifont}
\usepackage{enumerate}
\usepackage{xcolor}\definecolor{link}{RGB}{45,48,146}
\usepackage{hyperref}

\hypersetup{
pdftitle={Cavity Carving of Atomic Bell States (ArXiv Version)},
pdfauthor={Stephan Welte, Bastian Hacker, Severin Daiss, Stephan Ritter, and Gerhard Rempe},
colorlinks=true, linkcolor=link, citecolor=link, filecolor=link, urlcolor=link}

\newcommand{\unit}[1]{\ensuremath{\,\mathrm{#1}}}
\newcommand{\micro}{\ensuremath{\textnormal\textmu}}
\newcommand{\bra}[1]{{\langle{#1}|}}
\newcommand{\ket}[1]{{|{#1}\rangle}}
\newcommand{\braket}[1]{{\langle{#1}\rangle}}
\newcommand{\0}{\hphantom{0}}

\makeatletter
\@ifpackageloaded{hyperref}{\newcommand{\positionlabel}[1]{\hypertarget{#1}{}}}{\newcommand{\positionlabel}[1]{}}
\@ifpackageloaded{hyperref}{}{\newcommand{\hyperref}[2][]{#2}}
\@ifpackageloaded{hyperref}{\newcommand{\figref}[1]{\hyperlink{#1}{\ref*{#1}}}}{\newcommand{\figref}[1]{\ref{#1}}}
\makeatother

\begin{document}

\title{Cavity Carving of Atomic Bell States}
\author{Stephan~Welte}
\email{stephan.welte@mpq.mpg.de}
\author{\hspace{-.4em}\textcolor{link}{\normalfont\textsuperscript{$\dagger$}}\hspace{.4em}Bastian~Hacker}
\thanks{S.W.\ and B.H.\ contributed equally to this work.}
\author{Severin~Daiss}
\author{Stephan~Ritter}
\author{Gerhard~Rempe}
\affiliation{Max-Planck-Institut f\"ur Quantenoptik, Hans-Kopfermann-Strasse 1, 85748 Garching, Germany}

\begin{abstract}
We demonstrate entanglement generation of two neutral atoms trapped inside an optical cavity. Entanglement is created from initially separable two-atom states through carving with weak photon pulses reflected from the cavity. A polarization rotation of the photons heralds the entanglement. We show the successful implementation of two different protocols and the generation of all four Bell states with a maximum fidelity of $(90\pm2)\%$. The protocol works for any distance between cavity-coupled atoms, and no individual addressing is required. Our result constitutes an important step towards applications in quantum networks, e.g.\ for entanglement swapping in a quantum repeater.
\end{abstract}

\maketitle

Entanglement is a central ingredient of quantum physics. It was long debated until groundbreaking experiments with entangled photons \cite{Freedman1972,aspect1982}, ions \cite{turchette1998}, atoms \cite{hagley1997}, artificial atoms \cite{steffen2006}, and ensembles \cite{julsgaard2001} changed the view and launched an active field of research \cite{casabone2013, lin2013, kaufman2015}. Now entanglement is a powerful resource, with the teleportation of information in quantum networks among the most fascinating future applications \cite{kimble2008}. Elementary networks already exist, utilizing photons to distribute entanglement \cite{moehring2007, hofmann2012, ritter2012, bernien2013, delteil2016}. In such networks, cavity quantum electrodynamics (QED) not only ensures efficient connectivity between distant nodes \cite{reiserer2015} but also establishes enhanced capabilities. Most strikingly, network photons have been proposed as perfect workhorses to generate local entanglement \cite{soerensen2003a}.

Here we follow this proposal and show entanglement of two atomic qubits within one cavity QED node from which photons are reflected and detected with polarization-sensitive counters. As the scheme is insensitive to fluctuating photon numbers, we employ weak coherent laser pulses and use photon detection as a herald. Combined with atomic state rotations, we produce entangled states from an initially separable atom pair state, and show the creation of all four maximally entangled Bell states. Our scheme features several distinct advantages that distinguish it from other entanglement schemes \cite{lin2013, kaufman2015, wilk2010, isenhower2010}. Most notably, the interaction strength between two atoms coupled to the optical cavity does not depend on distance. Also, individual addressing of the two atoms is not required, rendering the technique robust, e.g., against focusing and pointing errors of the laser used for atomic state rotations. Moreover, the entangling protocol is fast, limited only by the duration of the atomic state rotation and the light pulses. The minimum pulse duration is determined by the cavity linewidth.

\begin{figure}[tb]
\centering
\positionlabel{fig:carvingexplanation}
\includegraphics[width=\columnwidth]{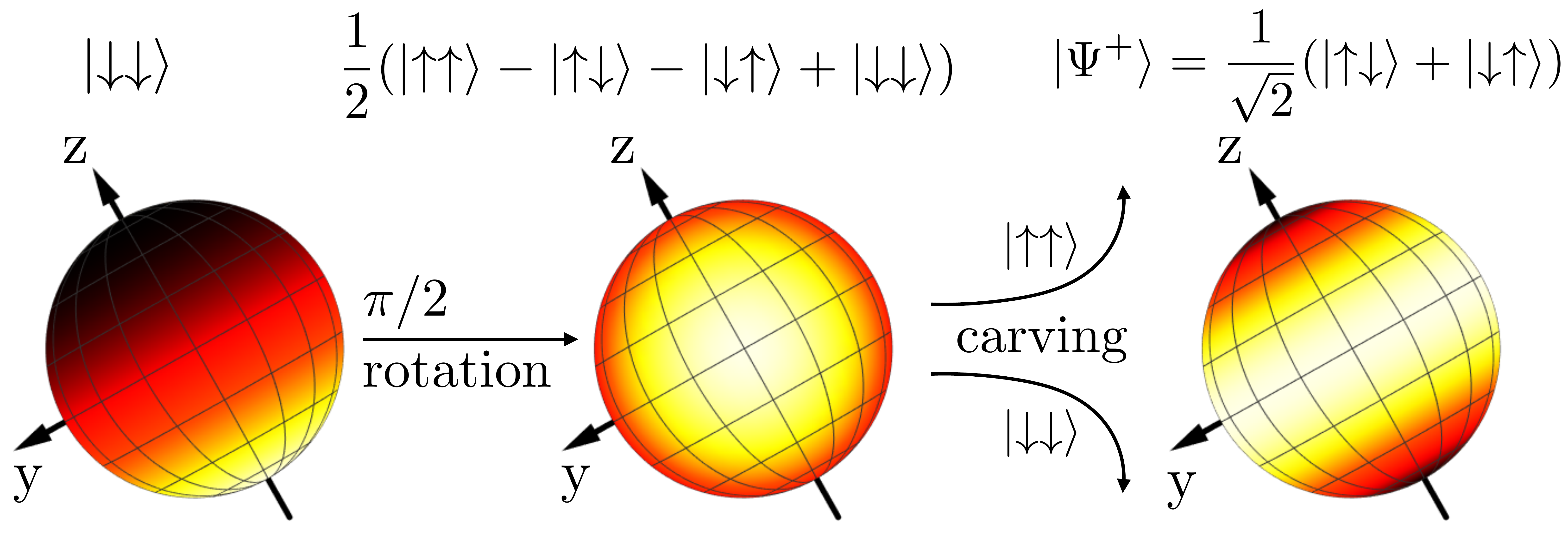}
\caption{\label{fig:carvingexplanation}
(color online). Illustration of quantum state car\-ving via the Husimi Q distribution. The Husimi Q distribution of a two-particle density matrix $\rho$ is defined as $Q(\rho;\theta,\phi)=\frac{3}{4\pi}\langle\theta,\phi\vert\rho\vert\theta,\phi\rangle$, where $\ket{\theta,\phi}=\scriptstyle\underset{j=1}{\overset{2}{\bigotimes}}\textstyle\bigl(\cos{(\theta/2)}\ket{{\uparrow}}_j-e^{i\phi}\sin{(\theta/2)}\ket{{\downarrow}}_j\bigr)$ are coherent spin states defined by the spherical coordinates $\theta$ and $\phi$ on the generalized Bloch sphere. The initial distribution (left sphere) is rotated around the $y$-axis (middle sphere) before the $\ket{{\downarrow}{\downarrow}}$ and $\ket{{\uparrow}{\uparrow}}$ components are carved out. After the carving step, the poles are not populated anymore and the maximally entangled Bell-state $\ket{\Psi^+}$ is prepared (right sphere). The color code is normalized on each sphere and $Q$ increases from dark to bright.}
\end{figure}

Following \cite{chen2015}, we call our technique carving. An initially separable state of two atoms undergoes a common projective measurement with probabilistic outcome. For an appropriately chosen projection subspace, the part of a two-atom wavefunction in that subspace can be entangled. If the measurement yields the orthogonal outcome, the atoms are not entangled and the attempt will be discarded. Specifically, each of the two atoms carries a qubit encoded in the states $\ket{{\uparrow}}$ and $\ket{{\downarrow}}$. Starting with an initially separable two-atom state $\ket{{\downarrow}{\downarrow}}$, a global $\pi/2$ rotation prepares $\frac{1}{2}(\ket{{\uparrow}{\uparrow}}-\ket{{\uparrow}{\downarrow}}-\ket{{\downarrow}{\uparrow}}+\ket{{\downarrow}{\downarrow}})$. A projective measurement (``carving'') allows us to probabilistically remove the $\ket{{\downarrow}{\downarrow}}$ and $\ket{{\uparrow}{\uparrow}}$ components of this state in a heralded protocol. In this way, we generate $\ket{\Psi^+}=\frac{1}{\sqrt{2}}(\ket{{\uparrow}{\downarrow}}+\ket{{\downarrow}{\uparrow}})$, a maximally entangled Bell state. The ideal carving process is illustrated with the Husimi Q distribution in Fig.\,\figref{fig:carvingexplanation}.

In our implementation, the projective measurement is performed by coherent light pulses, reflected off the cavity. We use linear polarization, $\ket{\text{A}}=\frac{1}{\sqrt2}(\ket{\text{L}}-i\ket{\text{R}})$, consisting of a right-circular component $\ket{\text{R}}$ which couples resonantly to any atom in $\ket{{\uparrow}}$ via the optical transition $\ket{{\uparrow}}\,{\leftrightarrow}\,\ket{e}$, and a left-circular component $\ket{\text{L}}$ as an uncoupled reference with the corresponding atomic transition far off-resonant. As both atoms are trapped in the same cavity mode, any atom in the coupling state $\ket{{\uparrow}}$ will change the reflection amplitude for $\ket{\text{R}}$. This leads to a nonzero probability to detect the reflected light in the orthogonal polarization mode $\ket{\text{D}}=\frac{1}{\sqrt2}(\ket{\text{L}}+i\ket{\text{R}})$. Therefore, any detection of a photon in $\ket{\text{D}}$ heralds the projection of the atoms into the subspace $\operatorname{span}(\{\ket{{\uparrow}{\uparrow}},\ket{{\uparrow}{\downarrow}},\ket{{\downarrow}{\uparrow}}\})$, and effectively carves away the $\ket{{\downarrow}{\downarrow}}$ component. Coherences within the subspace stay unaffected upon the projective measurement. The $\ket{\text{A}}\rightarrow\ket{\text{D}}$ flip probability $P_f=\bigl(\frac{\kappa_\text{out}}{\kappa}\frac{C}{C+1/2}\bigr)^2$ scales with the cooperativity $C$ (see \cite{supplement}). Here, $C=Ng^2/(2\kappa\gamma)$, with atom-cavity coupling rate $g$, number of coupling atoms $N$, total cavity field decay rate $\kappa$, decay rate through the outcoupling mirror $\kappa_\text{out}$, and atomic dipole decay rate $\gamma$ \cite{reiserer2015}. Our scheme requires neither strong coupling, $g>(\kappa,\gamma)$, nor high cooperativity, but both are beneficial for high efficiencies and high fidelities. In the case of high cooperativity and an asymmetric cavity, $C>\kappa_\text{out}/\kappa-1/2>0$ \cite{supplement}, as achieved in our experiment, any nonzero number of coupled atoms induces a $\pi$-phase shift on the reflected amplitude of $\ket{\text{R}}$ \cite{reiserer2014}. This makes the $\ket{\text{A}}\rightarrow\ket{\text{D}}$ polarization flip efficient even for one reflected photon. When more than one photon is reflected, the projected atomic state will not change further.

Compared to the proposed scheme \cite{soerensen2003a}, our version has the advantage that any light that is not matched to the geometric cavity mode will remain in its original polarization mode $\ket{\text{A}}$ and create no heralding signal in the $\ket{\text{D}}$ detector. This enhances the entangling fidelity significantly and makes the scheme robust against wavefront imperfections of the incident light.

\begin{figure}[tb]
\centering
\positionlabel{fig:setup}
\includegraphics[width=\columnwidth]{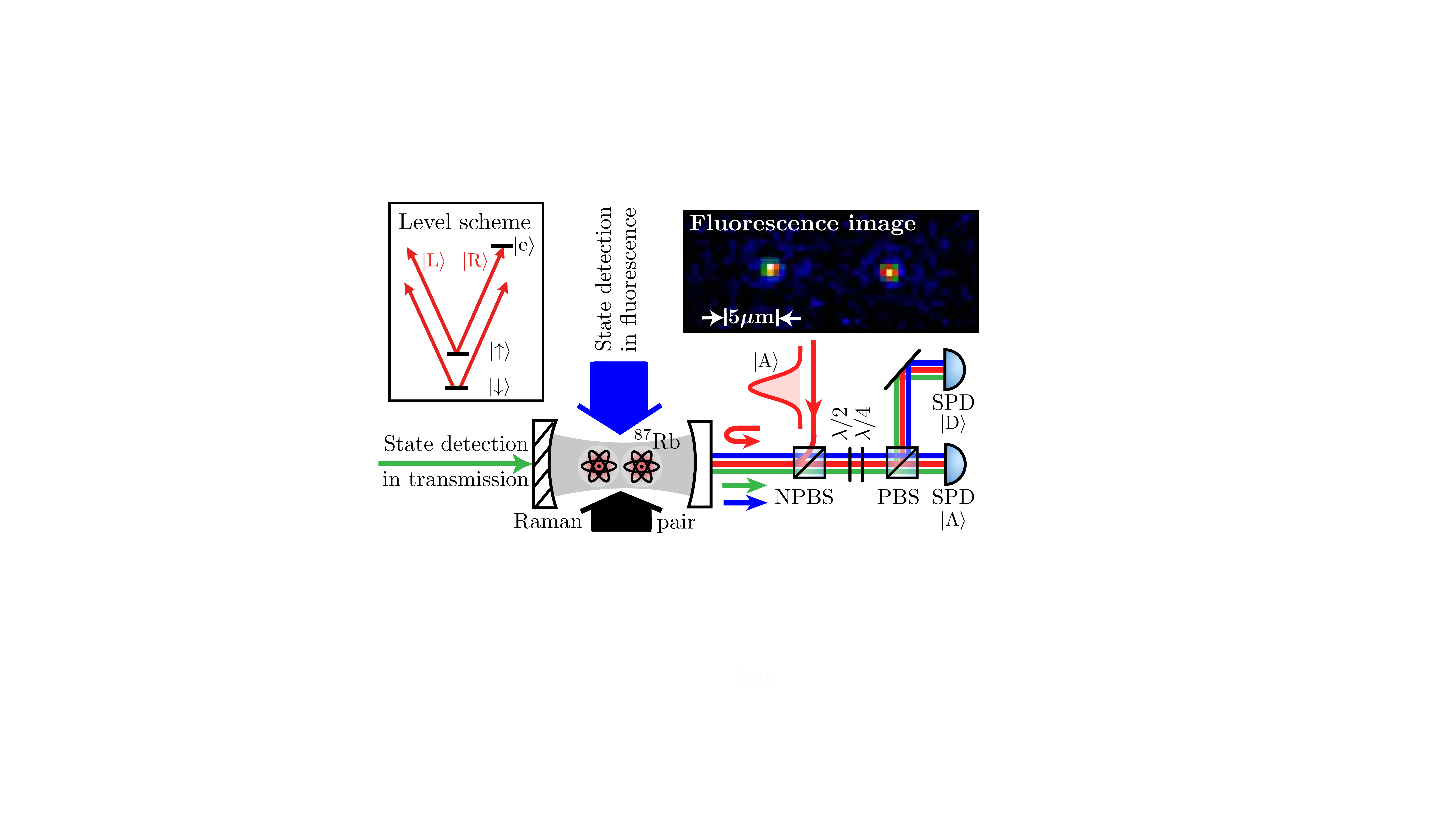}
\caption{\label{fig:setup}
(color online). Schematic representation of the setup.
Two \textsuperscript{87}Rb atoms are trapped near the cavity center. The internal state of the atoms is controlled with a Raman laser pair impinging from the side of the cavity, colinear with a state-detection beam (blue). State detection of the atomic qubits can additionally be performed with a laser probing the cavity transmission (green). Coherent light pulses impinge on the cavity's outcoupling mirror through a non-polarizing beam splitter (NPBS) and are reflected from the cavity. Light from the cavity is detected by a polarization-resolving detection setup consisting of a half- and a quarter-wave plate ($\lambda/2$ and $\lambda/4$), a polarizing beam splitter (PBS) and two single-photon detectors (SPD). The left inset shows a simplified level scheme of a single atom with the two ground states $\ket{{\uparrow}}$ and $\ket{{\downarrow}}$ and the excited state $\ket{\text{e}}$. $\ket{\text{R}}$ photons couple $\ket{{\uparrow}}$ and $\ket{\text{e}}$. The right inset shows a fluorescence image of two trapped atoms.}
\end{figure}

The experimental setup is an extension of earlier work \cite{reiserer2014} which now allows to trap two \textsuperscript{87}Rb atoms in a three-dimensional optical lattice \cite{supplement} at the cavity center with trapping times of several seconds. The largest separation $d$ between the atoms is along an axis perpendicular to the cavity. Via fluorescence images (right inset in Fig.\,\figref{fig:setup}), we ensure that only atom pairs positioned symmetrically around the cavity mode center, with a distance $2\unit{{\micro}m}\le{d}\le12\unit{{\micro}m}$, are being used. A blue-detuned optical lattice along the cavity axis confines the atoms close to anti-nodes of the $780\unit{nm}$ cavity field and thereby ensures coupling to the cavity on each trapping site.

The cavity with length $486\unit{{\micro}m}$ and mode waist $30\unit{{\micro}m}$ is single-sided with mirror transmissions of $4.0{\times}10^{-6}$ and $9.2{\times}10^{-5}$, allowing for an efficient in- and outcoupling of light in one direction. It is actively tuned into resonance with the atomic transition $\ket{{\uparrow}}:=\ket{F{=}2,m_F{=}2}\leftrightarrow\ket{e}:=\ket{F'{=}3,m_F{=}3}$ on the D$_2$ line. On this transition, we achieve $(g,\kappa,\kappa_\text{out},\gamma)=2\pi\,(7.8,2.5,2.3,3.0)\unit{MHz}$ and thus a cooperativity $C=4.1$ for one coupling atom.

We choose the ground state $\ket{{\downarrow}}:=\ket{F{=}1,m_F{=}1}$ as our second qubit state (left inset in Fig.\,\figref{fig:setup}). For each experiment, we employ an experimental sequence of state preparation, quantum state carving, analysis and readout \cite{supplement}. Depending on the desired entangled output state, we start by preparing the atoms in $\ket{{\downarrow}{\downarrow}}$ or in a statistical mixture of anti-parallel states with density matrix $\frac12\ket{{\uparrow}{\downarrow}}\bra{{\uparrow}{\downarrow}}+\frac12\ket{{\downarrow}{\uparrow}}\bra{{\downarrow}{\uparrow}}$. These initial states can be realized by means of a dynamical Stark detuning of the optical transitions induced by the power of the trap laser \cite{supplement}. Coherent qubit control is achieved with a pair of Raman lasers which copropagate perpendicular to the cavity axis and illuminate both atoms equally with a beam waist $w_0=35\unit{{\micro}m}$, much bigger than the inter-atomic distance. We denote rotations as ``$R$'' with the rotation angle as a superscript, and with a subscript defining the rotation axis $x$ ($\ket{\uparrow}+\ket{\downarrow}$), $y$ ($\ket{\uparrow}+i\ket{\downarrow}$), or $z$ ($\ket{\uparrow}$). State rotations from the initial state $\ket{{\downarrow}{\downarrow}}$ create coherent spin states $\ket{\theta,\phi}$ where $\theta$ and $\phi$ can be controlled via the Raman laser power, duration, detuning and phase.

We employ a state-detection protocol consisting of two successive measurements on the two atoms with an interleaved $\pi$ pulse. This allows us to discriminate between $\ket{{\downarrow}{\downarrow}}$, $\ket{{\uparrow}{\uparrow}}$ and $\ket{{\uparrow}{\downarrow}}$/$\ket{{\downarrow}{\uparrow}}$ \cite{supplement}.

The coherent light pulses for carving ($0.7\unit{{\micro}s}$ full-width at half maximum) impinge on the cavity, are reflected, and measured polarization-resolved with separate single-photon detectors for $\ket{\text{A}}$ and $\ket{\text{D}}$ (Fig.\,\figref{fig:setup}).

\begin{figure}[t]
\centering
\positionlabel{fig:circuitdiagram}
\includegraphics[width=\columnwidth]{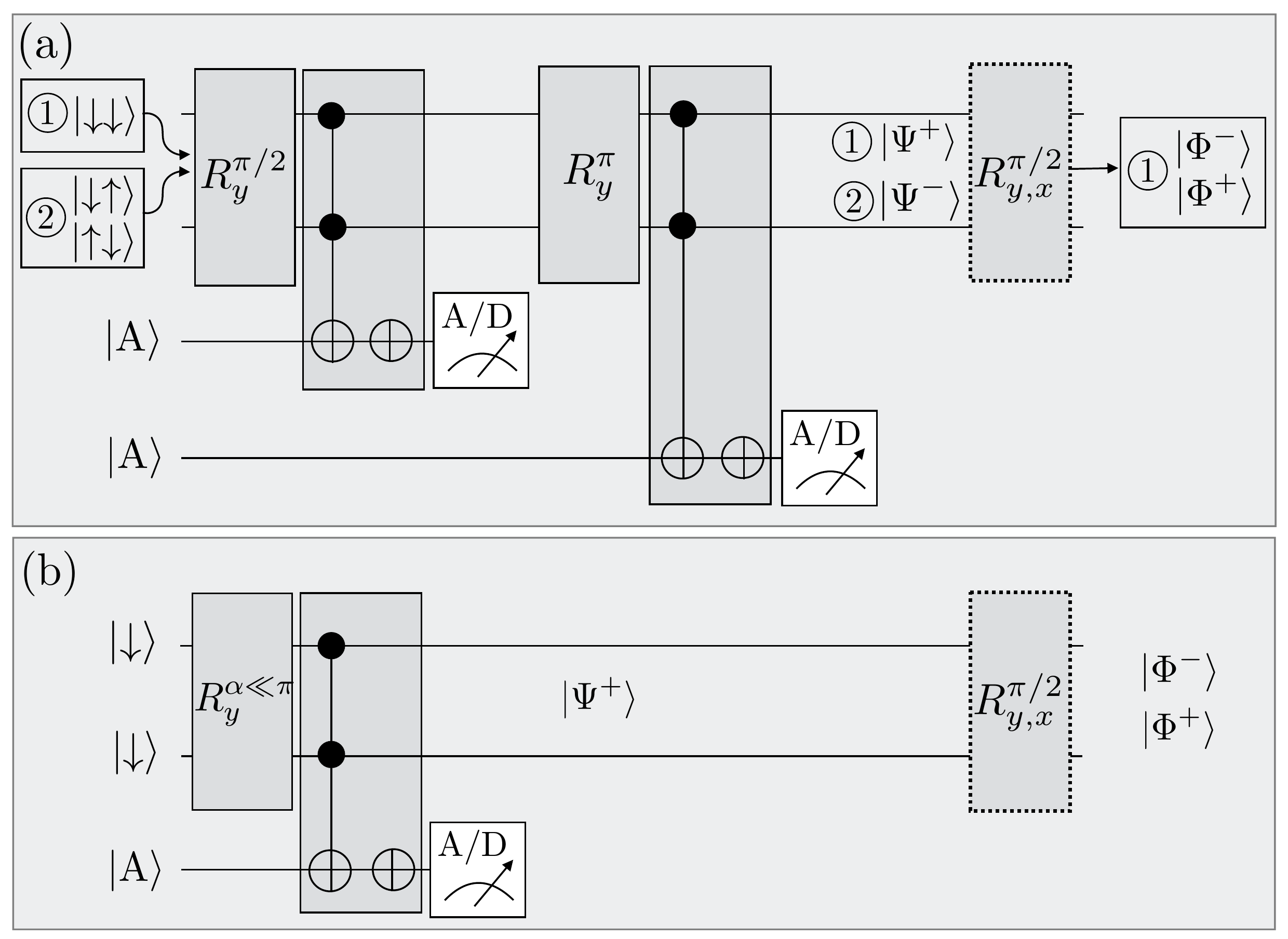}
\caption{\label{fig:circuitdiagram}
Quantum circuit diagrams.
(a) Double-carving scheme. If the initial atom-atom state is chosen as \textcircled{1} $\ket{{\downarrow}{\downarrow}}$, the symmetric state $\ket{\Psi^+}$ is prepared via the reflection of two antidiagonally ($\ket{\text{A}}$) polarized photons, each followed by postselection, and interleaved by a $\pi$ pulse. A subsequent, optional $\pi/2$ rotation (dashed box) generates $\ket{\Phi^\pm}$ depending on the phase of the last rotation pulse. Starting with atoms in opposite states \textcircled{2} $\ket{{\uparrow}{\downarrow}}$ or $\ket{{\downarrow}{\uparrow}}$ or a statistical mixture of them, the singlet state $\ket{\Psi^-}$ can be prepared with the same protocol.
(b) Single-carving scheme following weak initial excitation.  Starting from $\ket{{\downarrow}{\downarrow}}$, the scheme creates $\ket{\Psi^+}$ which can afterwards be transformed into $\ket{\Phi^\pm}$ by an optional $\pi/2$-pulse (dashed box) of appropriate phase. The state $\ket{\Psi^-}$ can not be created with this scheme.}
\end{figure}
We demonstrate the carving technique in a scheme called ``double carving'' which is adapted from Ref.\,\cite{soerensen2003a} [Fig.\,\figref{fig:circuitdiagram}(a)].  We start by preparing the separable state $\ket{{\downarrow}{\downarrow}}$, from which $\frac{1}{2}(\ket{{\uparrow}{\uparrow}}-\ket{{\uparrow}{\downarrow}}-\ket{{\downarrow}{\uparrow}}+\ket{{\downarrow}{\downarrow}})$ is created via a global $R^{\pi/2}_y$ rotation. Our default photon number per reflection pulse is $\overline{n}=0.33$. The first reflection of an $\ket{\text{A}}$-polarized pulse is followed by the projection of the atomic state to $\frac{1}{\sqrt{3}}(\ket{{\uparrow}{\uparrow}}-\ket{{\uparrow}{\downarrow}}-\ket{{\downarrow}{\uparrow}})$ whenever photons are detected in $\ket{\text{D}}$, which happens in 61\% of the detection events (3/4 for an ideal cavity). Then, we apply a global $R^\pi_y$ rotation yielding $\frac{1}{\sqrt{3}}(\ket{{\downarrow}{\downarrow}}+\ket{{\uparrow}{\downarrow}}+\ket{{\downarrow}{\uparrow}})$ and reflect a second $\ket{\text{A}}$-polarized pulse. It is detected in the orthogonal state with 53\% probability (2/3 for an ideal cavity), resulting in $\ket{\Psi^+}=\frac{1}{\sqrt{2}}(\ket{{\uparrow}{\downarrow}}+\ket{{\downarrow}{\uparrow}})$. Effectively, this protocol removes the populations of $\ket{{\uparrow}{\uparrow}}$ and $\ket{{\downarrow}{\downarrow}}$ from the initially rotated state, yielding a total success probability of $32\%$ ($1/2$ for an ideal cavity), as only runs where the polarization of both photons flipped are postselected. With an average detected 0.11 photons per pulse, the experiment succeeds with a total efficiency of $(0.38\pm0.01)\%$.

\begin{figure}[tb]
\centering
\positionlabel{fig:parity}
\includegraphics[width=\columnwidth]{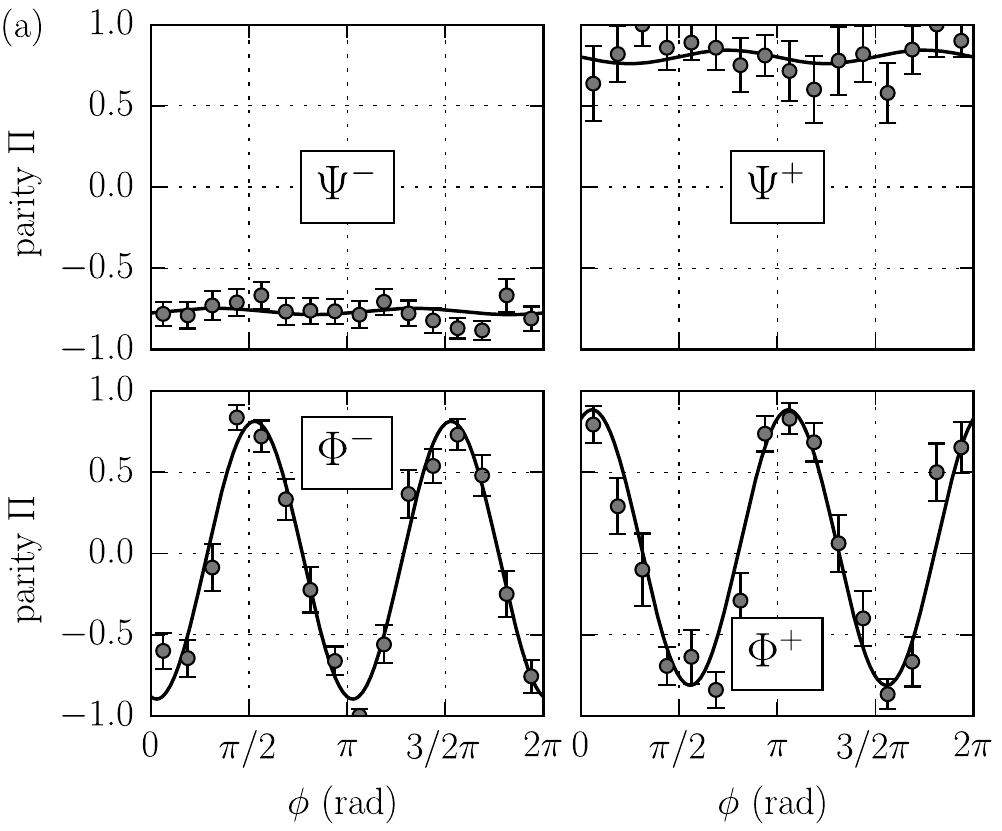}
\includegraphics[width=\columnwidth]{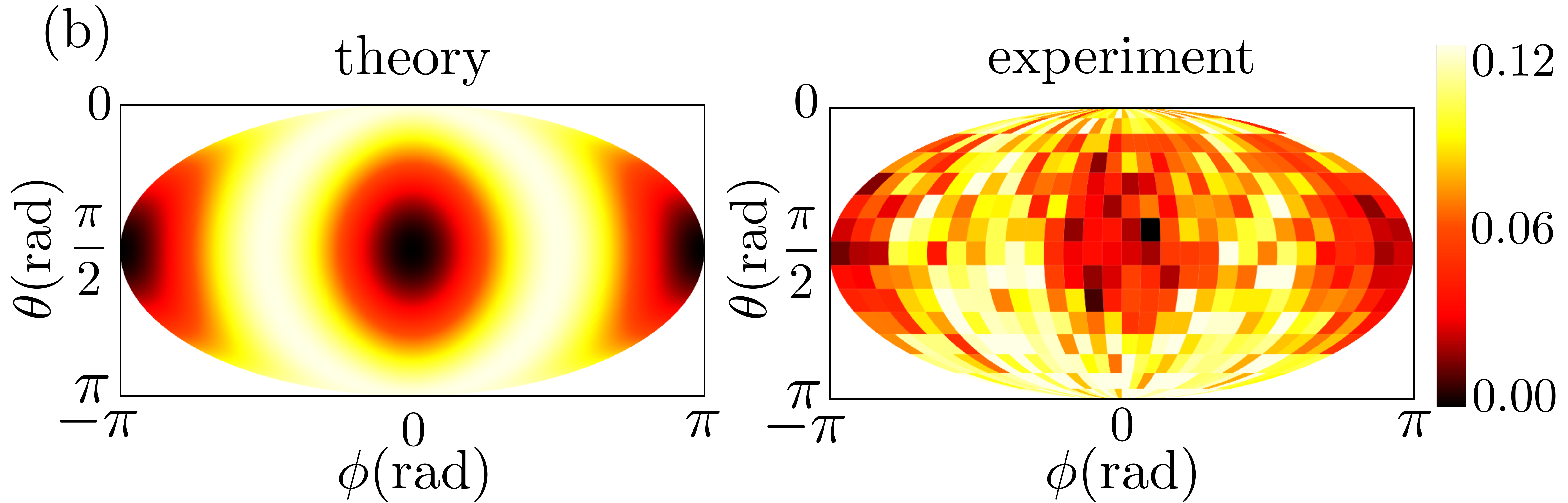}
\caption{\label{fig:parity}
(color online). Parity oscillations of the four Bell states and Husimi Q distribution of $\ket{\Phi^-}$.
(a) Shown are experimental data for $\Pi(\phi)$ from the double-carving entangling scheme with standard errors of each mean value, and a fit of $\Pi(\phi)$ with $\operatorname{Re}(\rho_{{\uparrow}{\downarrow},{\downarrow}{\uparrow}})$, $\operatorname{Im}(\rho_{{\uparrow}{\uparrow},{\downarrow}{\downarrow}})$ and $\operatorname{Re}(\rho_{{\uparrow}{\uparrow},{\downarrow}{\downarrow}})$ as free parameters.
(b) Comparison between the theoretically expected Husimi distribution and experimental data for the $\ket{\Phi^-}$ state.  The Mollweide projection of the respective spheres is shown.}
\end{figure}

To analyze a generated two-atom output state with density matrix $\rho$, we determine its fidelity $F{=}\braket{\psi\,\vline\,\rho\,\vline\,\psi}$ with respect to an ideal state $\ket{\psi}$. This requires a direct measurement of the state's populations $P_{{\uparrow}{\uparrow}}$, $P_{{\downarrow}{\downarrow}}$ and $(P_{{\uparrow}{\downarrow}}+P_{{\downarrow}{\uparrow}})$ and the determination of coherences \cite{supplement}. To retrieve the latter, we use the method of parity oscillations \cite{turchette1998,sackett2000}: An additional analysis pulse of area $\pi/2$ is inserted right before the state detection. The rotation axis, given by the phase $\phi$ with respect to the preparation Raman-beam pulses, is scanned from $0$ to $2\pi$ over 750 subsequent experiments and we measure the resulting populations $\tilde P(\phi)$. Experiments are repeated at a rate of $1\unit{kHz}$ with $180\unit{{\micro}s}$ being used for optical pumping and $740\unit{{\micro}s}$ for cooling between each experiment. The parity $\Pi(\phi):=\tilde P_{{\uparrow}{\uparrow}}+\tilde P_{{\downarrow}{\downarrow}}-\tilde P_{{\uparrow}{\downarrow}}-\tilde P_{{\downarrow}{\uparrow}}$ oscillates as $\Pi(\phi)=2\operatorname{Re}(\rho_{{\uparrow}{\downarrow},{\downarrow}{\uparrow}})+2\operatorname{Im}(\rho_{{\uparrow}{\uparrow},{\downarrow}{\downarrow}})\sin(2\phi)+2\operatorname{Re}(\rho_{{\uparrow}{\uparrow},{\downarrow}{\downarrow}})\cos(2\phi)$ (Fig.\,\figref{fig:parity}) and yields information about coherences through a fit of the oscillation amplitude, phase and offset. Here, $\rho_{ij,kl}$ denote elements of the density matrix $\rho$ with $i,j,k,l\in\{{\uparrow},{\downarrow}\}$.

The generated $\ket{\Psi^+}$ state shows a high offset value of $2\operatorname{Re}(\rho_{{\uparrow}{\downarrow},{\downarrow}{\uparrow}})=(80\pm5)\%$ (Fig.\,\figref{fig:parity} upper right). The fidelity with the expected Bell state is calculated according to $F(\Psi^+)=\frac12(P_{{\uparrow}{\downarrow}}+P_{{\downarrow}{\uparrow}})+\operatorname{Re}(\rho_{{\uparrow}{\downarrow},{\downarrow}{\uparrow}})=(81.9\pm2.8)\%$. A fidelity above $50\%$ with any Bell state proves entanglement.

The $\ket{\Psi^+}$ state is one of the Bell triplet states, and can be transformed into $\ket{\Phi^-}=\frac{1}{\sqrt{2}}(\ket{{\uparrow}{\uparrow}}-\ket{{\downarrow}{\downarrow}})$ via a  $R^{\pi/2}_y$ rotation or into $\ket{\Phi^+}=\frac{1}{\sqrt{2}}(\ket{{\uparrow}{\uparrow}}+\ket{{\downarrow}{\downarrow}})$ through a $R^{\pi/2}_x $ rotation (Fig.\,\figref{fig:circuitdiagram}). Both of them exhibit the expected oscillations. Lastly, we can create the Bell singlet state $\ket{\Psi^-}=\frac{1}{\sqrt{2}}(\ket{{\uparrow}{\downarrow}}-\ket{{\downarrow}{\uparrow}})$, using the same carving protocol, but starting with an initial density matrix $\frac12\ket{{\uparrow}{\downarrow}}\bra{{\uparrow}{\downarrow}}+\frac12\ket{{\downarrow}{\uparrow}}\bra{{\downarrow}{\uparrow}}$. A rotation by $\pi/2$ creates coherent $\ket{{\downarrow}{\uparrow}}$ and $\ket{{\uparrow}{\downarrow}}$ contributions with opposite sign, which remain coherently preserved throughout the rest of the protocol. As both $\ket{{\downarrow}{\downarrow}}$ and $\ket{{\uparrow}{\uparrow}}$ contributions are carved out, we arrive at a pure $\ket{\Psi^-}$. The parity data for all four states is shown in Fig.\,\figref{fig:parity}, which additionally displays the measured Husimi Q distribution for an experimentally prepared Bell state $\ket{\Phi^-}$. Table \ref{tab:fidelities} lists the measured populations and the fidelities for all four Bell states. Furthermore, we investigate the lifetimes $\tau$ of the entangled states by measuring the fidelities after various waiting intervals. The $1/e$ values of a fitted Gaussian are given in Tab.\,\ref{tab:fidelities}. We find $\tau$ to be on the order of a few hundred microseconds, limited by fluctuating real and virtual, i.e.\ trap-induced, magnetic fields of a few mG and mechanical-state-dependent dynamical Stark shifts. The $\ket{\Psi^\pm}$ states live longer than the $\ket{\Phi^-}$ state, in accordance with the expectation that they are insensitive to common-mode field fluctuations \cite{Lidar1998}.
\begin{table}[b]
\caption{\label{tab:fidelities}
Measured populations $P$, fidelities $F$ and lifetimes $\tau$ of states in our experiment.}
\begin{ruledtabular}
\begin{tabular}{lccccc}
$\ket{\psi}$&$P_{{\uparrow}{\uparrow}}$&$P_{{\downarrow}{\downarrow}}$&$P_{{\uparrow}{\downarrow}}+P_{{\downarrow}{\uparrow}}$&$F$&$\tau\;(\mathrm{{\micro}s})$\\
\hline
$\ket{\Psi^-}$&\06(2)\%&\09(2)\%&84(2)\%&$83.4(1.4)\%$&$204(26)$\\
$\ket{\Psi^+}$&\02(2)\%&15(5)\%&83(5)\%&$81.9(2.8)\%$&$134(17)$\\
$\ket{\Phi^-}$&40(3)\%&54(3)\%&\06(1)\%&$89.9(1.7)\%$&$\090(19)$\\
$\ket{\Phi^+}$&44(5)\%&43(5)\%&13(4)\%&$82.4(3.1)\%$&---
\end{tabular}
\end{ruledtabular}
\end{table}

\begin{figure}[tb]
\positionlabel{fig:Fvsangle}
\includegraphics[width=0.5\columnwidth]{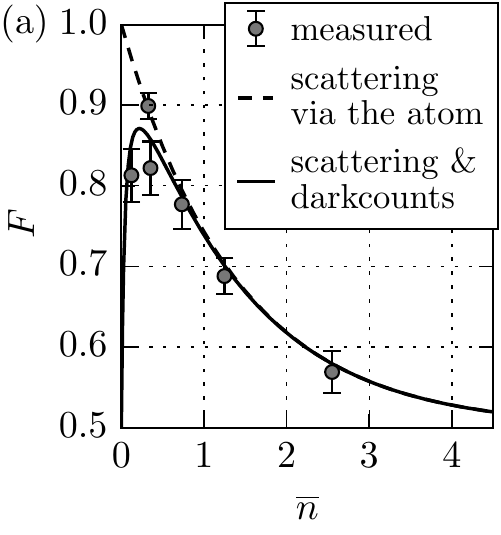}
\nolinebreak
\includegraphics[width=0.5\columnwidth]{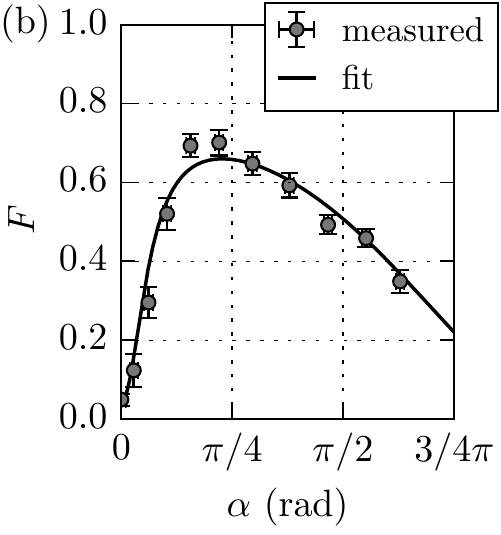}
\caption{\label{fig:Fvsangle}
(a) Entangled state fidelity $F$ vs.\ average incident photon number $\overline{n}$ in each coherent pulse in the double-carving scheme. The dashed line shows the simple theoretical model of exponential decoherence through photons scattered by the atoms. The predicted decay rate for our cavity parameters is applied without any free fitting parameters. The model plotted as a solid line additionally considers dark counts, which become dominant for low $\overline{n}$, and shows a fitted dark count rate of 0.011 per pulse.
(b) Fidelity vs.\ rotation angle $\alpha$ in the single-carving scheme, showing measured data and a fitted model including dark counts. All error bars are standard deviations of the mean.}
\end{figure}

According to the model of \cite{soerensen2003a}, scattering of carving photons by atomic spontaneous decay leads to decoherence of the entangled state. With an impinging coherent pulse of $\overline{n}$ photons on average, the expected number of scattered photons is given by $\overline{n}\cdot{s}$, with scattering fraction $s$, independent of the number of un-scattered detected photons in each shot. With our cavity parameters we have $s=4\kappa_\text{out}\gamma{N}g^2/(\kappa\gamma+Ng^2)^2=0.36$ for $N=1$. With two reflected pulses, twice the amount of decoherence is expected. We compare this model to measured values of the fidelity for various $\overline{n}$, and find very good agreement [Fig.\,\figref{fig:Fvsangle}(a)]. Only for very low $\overline{n}$ the measured fidelities drop below this simple model, because detector dark counts become important. Thus, we conclude that spontaneous decay and detector dark counts are the dominant processes that deteriorate $F$, and that an improvement would require lower $s$, for example through an increased atom-cavity coupling rate $g$.

We realize a second entangling protocol, named ``single-carving'' scheme [Fig.\,\figref{fig:circuitdiagram}(b)] and also proposed in Ref.\,\cite{soerensen2003a}. It employs a small $R^\alpha_y$ rotation with \mbox{$\alpha\ll\pi$} on the initial $\ket{{\downarrow}{\downarrow}}$ state followed by one carving step. This rotation creates a state with only a small $\ket{{\uparrow}{\uparrow}}$ contribution
\begin{align}
\label{eq:singlecarvingprep}
\sin^2\frac\alpha2\,\ket{{\uparrow}{\uparrow}}&-\frac12\sin\alpha\,(\ket{{\uparrow}{\downarrow}}+\ket{{\downarrow}{\uparrow}})+\cos^2\frac\alpha2\,\ket{{\downarrow}{\downarrow}}
,
\end{align}
which can be converted into a highly entangled state by only one reflection-based carving of $\ket{{\downarrow}{\downarrow}}$. The resulting state is approximately a $\ket{\Psi^+}$ state for small $\alpha$ and can afterwards be rotated to a $\ket{\Phi^-}$. The scheme has an intrinsic trade-off between achievable efficiency $\eta=1-\cos^4(\alpha/2)$ and fidelity $F=4\cos^2(\alpha/2)\mathbin{/}(3+\cos\alpha)$ with $\alpha$ as an adjustable parameter additionally to the reflection pulse photon number $\overline{n}$. In our experiment we create the $\ket{\Phi^-}$ state with a relatively large $\overline{n}=1.2$, and varied $\alpha$ between 0 and $0.63\,\pi$ [Fig.\,\figref{fig:Fvsangle}(b)]. For large $\alpha\gg0$ we observe the expected drop of the fidelity as the  undesired component in Eq.(\ref{eq:singlecarvingprep}) increases. For very small $\alpha$, $F$ decreases again, as the signal becomes quickly dominated by detector dark counts of 0.01 per pulse. For $\alpha=0.23\,\pi$ we find a maximum fidelity of $F(\Phi^-)=(70.2\pm3.2)\%$. This fidelity is limited by both dark counts and $\overline{n}$.  With our chosen settings we reach a heralding efficiency of $(5.9\pm0.1)\%$, benefiting from the larger $\overline{n}$ compared to the double-carving scheme. This second scheme can directly be extended to three or more atoms. The feasibility to entangle atomic ensembles has been shown in \cite{haas2014,mcconnell2015}.

To compare, we found the double-carving scheme to yield higher entangling fidelities, limited only by photon scattering and detector dark counts. The efficiency is larger in our implementation of the single-carving scheme. In the presence of dark counts, the choice between different protocols and parameters therefore offers a convenient way to optimize between experimentally achievable efficiencies and fidelities.

Our cavity can readily act as a quantum network node \cite{ritter2012} and local entanglement between two atoms can be mapped onto photons and then be distributed in the network. In a quantum repeater scheme based on cavity QED systems \cite{uphoff2016}, the presented double-carving scheme combined with a third photon reflection can furthermore serve as an entanglement-swapping protocol. The straightforward extensions towards more atoms provides a promising route to explore the full scalability of the system.

\begin{acknowledgments}
We thank M.\ K\"orber, A.\ Neuzner, A.\ Reiserer and M.\ Uphoff  for valuable ideas and discussions. This work was supported by the Bundesministerium f\"ur Bildung und Forschung via IKT 2020 (Q.com-Q) and by the Deutsche Forschungsgemeinschaft via the excellence cluster Nanosystems Initiative Munich (NIM). S.W.\ was supported by the doctorate programme Exploring Quantum Matter (ExQM).
\end{acknowledgments}

\nocite{walls2008, neuzner2016, reimann2015}

\clearpage

\section{SUPPLEMENTAL MATERIAL}
\label{supplement}

\subsection{Required Cavity Parameters}
While the carving scheme proposed by S{\o}rensen and M{\o}lmer \cite{soerensen2003a_sup} works well only in the case of an exactly symmetric cavity with $\kappa_\text{out}=\kappa/2$, it can be extended to arbitrary combinations of outcoupling rate $\kappa_\text{out}$ and total cavity field decay rate $\kappa$. To this end, we use two polarization modes, $\ket{\text{R}}$ coupling to the atoms and $\ket{\text{L}}$ far off-resonant serving as a reference. Then, instead of heralding on any reflected photon, we impinge linear photons $\ket{\text{A}}=\frac{1}{\sqrt2}(\ket{\text{L}}-i\ket{\text{R}})$ and look for photons of orthogonal polarization with respect to the incident mode. This modification makes the scheme not only robust against variations of the cavity QED parameters, it also lifts the requirement of perfect transversal optical mode matching between the incoming beam and the cavity, as unmatched light is reflected without polarization flip and triggers no heralding signal. This is crucial since it is challenging to achieve a mode matching above 90\% in our experiment.

We now derive the  success rate of our scheme which depends on the cavity QED parameters $g$, $\kappa$, $\kappa_\text{out}$ and $\gamma$. The number of coupling atoms is denoted by $N$. For our cavity, these parameters are $(g,\kappa,\kappa_\text{out},\gamma)=2\pi\,(7.8,2.5,2.3,3.0)\unit{MHz}$ making the cooperativity $C=Ng^2/(2\kappa\gamma)=4.1$ for $N=1$ coupling atom. In the limit of long photonic wavepackets (compared to the cavity decay time), the reflection amplitude $r$ of resonant light impinging on the outcoupling mirror is given by \cite{walls2008_sup}
\begin{equation}
r(N)=1-\frac{2\kappa_\text{out}\gamma}{Ng^2+\kappa\gamma}=1-\frac{\kappa_\text{out}/\kappa}{C+1/2}\ .
\end{equation}
The reflection amplitude with $N$ coupling atoms holds for $\ket{\text{R}}$-light, whereas $\ket{\text{L}}$ always gets the amplitude $r(0)$. The reflection operator, expressed in the $\ket{\text{R}}$/$\ket{\text{L}}$-basis is diagonal: $\hat{R}=\ket{\text{R}}\bra{\text{R}}r(N)+\ket{\text{L}}\bra{\text{L}}r(0)$. When we impinge a linear input polarization $\ket{\text{A}}=\frac{1}{\sqrt2}(\ket{\text{L}}-i\ket{\text{R}})$, the probability for the photon being reflected in the orthogonal state $\ket{\text{D}}=\frac{1}{\sqrt2}(\ket{\text{L}}+i\ket{\text{R}})$ is
\begin{align}
P_f
&=\left|\bra{\text{D}}\hat{R}\ket{\text{A}}\right|^2=\nonumber\\
&=\left|\textstyle\frac12r(N{=}0)-\textstyle\frac12r(N{>}0)\right|^2=\nonumber\\
&=\left(\frac{\kappa_\text{out}}{\kappa}\frac{Ng^2}{Ng^2+\gamma\kappa}\right)^2
=\left(\frac{\kappa_\text{out}}{\kappa}\frac{C}{C+1/2}\right)^2
\end{align}
and becomes non-zero whenever $N>0$ and $g>0$.

The total success probability depends on the number of heralding photons $\overline{n}\,P_f$. This favors large $\overline{n}$. However, when the experiment is performed with coherent light pulses containing several photons, the achievable fidelity drops with the number of scattered photons $\overline{n}\,s$ as described in the main text. This favors small $\overline{n}$. Best performance is achieved for a large ratio $P_f/s=\frac{Ng^2\kappa_{\text{out}}}{4\kappa^2\gamma}=\frac{\kappa_\text{out}}{2\kappa}C$, with the cooperativity as the key parameter and an asymmetric cavity with $\kappa_\text{out}\approx\kappa$.

The carving scheme becomes efficient when most of the reflected photons flip their linear polarization for $N>0$. On resonance, this is the case when the reflection amplitude $r(N{>}0)$ for right-circularly polarized coupling light changes its sign, i.e.\ the atoms produce a $\pi$-phase-shift in $r$. The phase of the reflected light is given by $\arg(r)$, the angle of $r$ with the positive real axis. The above expression for $r$ on resonance shows that $r(N{>}0)>r(N{=}0)$. This means that the following conditions need to be fulfilled for the $\pi$-phase shift: 
\begin{align}
r(N{>}0)&=1-\frac{2\kappa_\text{out}\gamma}{Ng^2+\kappa\gamma}=1-\frac{\kappa_\text{out}/\kappa}{C+1/2}>0\\
r(N{=}0)&=1-\frac{2\kappa_\text{out}}{\kappa}<0
\end{align}
Here the first inequality $N g^2>\gamma(2\kappa_\text{out}-\kappa)$ is a condition for high cooperativity and the second condition $\kappa_\text{out}>\kappa/2$ implies an asymmetric cavity. Within these limits, the polarization flip of the reflected photons can be expressed as a truthtable that heralds the existence of coupled atoms:
\begin{align}
\label{eq:gateD}
&\ket{{\uparrow}{\uparrow}\,\text{A}}\ \rightarrow\ \ket{{\uparrow}{\uparrow}\,\text{D}}\nonumber\\
&\ket{{\uparrow}{\downarrow}\,\text{A}}\ \rightarrow\ \ket{{\uparrow}{\downarrow}\,\text{D}}\nonumber\\
&\ket{{\downarrow}{\uparrow}\,\text{A}}\ \rightarrow\ \ket{{\downarrow}{\uparrow}\,\text{D}}\nonumber\\
&\ket{{\downarrow}{\downarrow}\,\text{A}}\ \rightarrow\ \ket{{\downarrow}{\downarrow}\,\text{A}}
\end{align}
The total reflectivity on resonance in our system $|r(0)|^2=0.71$, $|r(1)|^2=0.64$ and $|r(2)|^2=0.80$ depends only slightly on the number of coupling atoms.

\subsection{Qubit Preparation and Readout}
\label{sec:stateprepdet}
The full experimental sequence of the double-detection carving of Bell states is shown as a quantum circuit diagram in Fig.\,\figref{fig:stateprep}. All operations for the preparation, manipulation and detection of the state of the two atoms are applied globally, with beams addressing both atoms equally. Even though a discrimination between the two atoms is in principle possible due to their spatial separation, it is not required here. This simplifies the entangling operation from a technological point of view and makes it more robust, e.g.\ against variations in the separation of the atoms or beam pointing instabilities.

\renewcommand{\thefigure}{S1}
\begin{figure}[h]
\centering
\positionlabel{fig:stateprep}
\includegraphics[width=1.0\columnwidth]{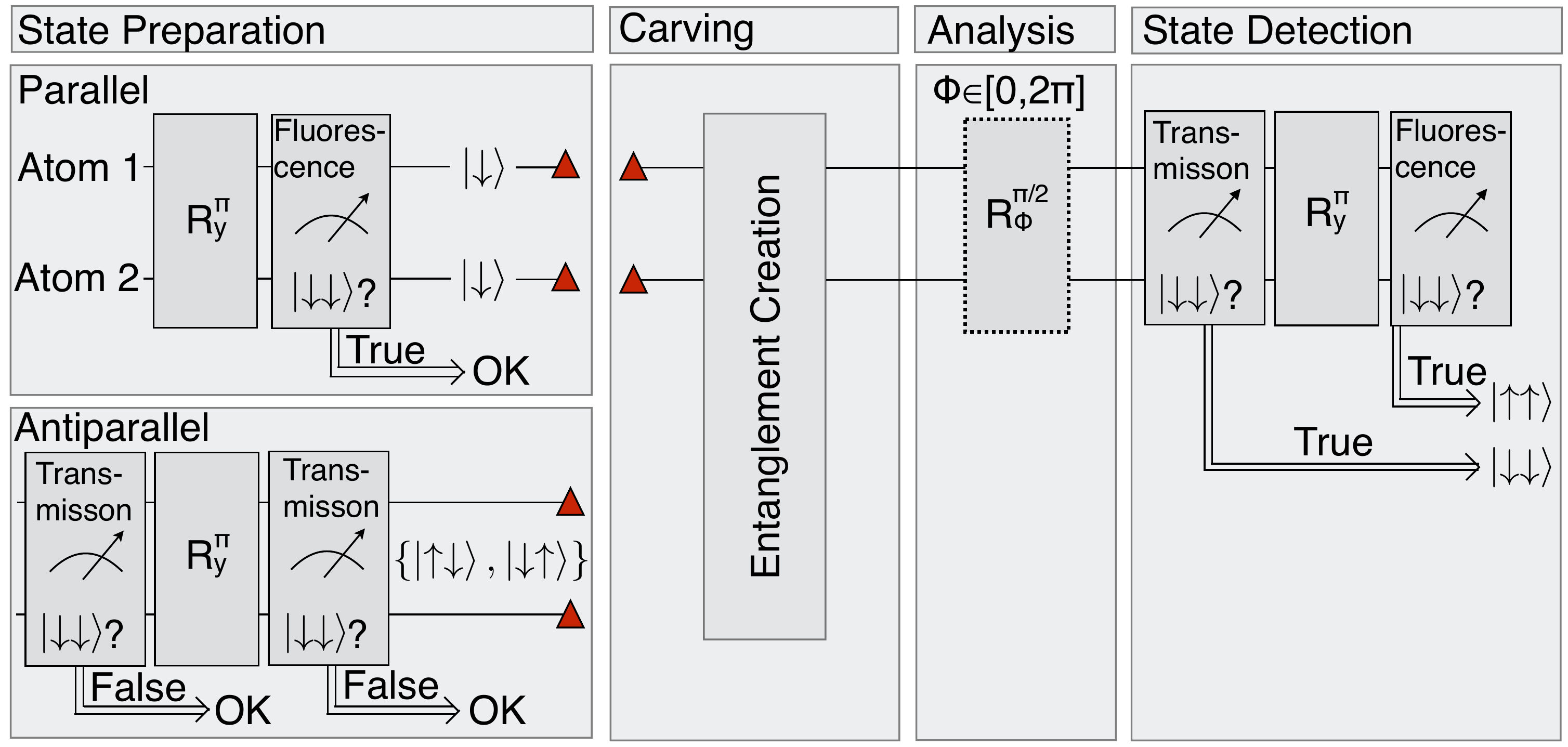}
\caption{\label{fig:stateprep}
(color online). The entanglement sequence described in the main text is preceded by a preparation of the two atomic qubits and is followed by a readout procedure to characterize the created two-particle state. Initial states are prepared by optical pumping. Employing one of the two state-preparation protocols, parallel or antiparallel states can be prepared experimentally. After the entangling experiment, states may be rotated with an analysis pulse of varying phase, and are read out in a two-step protocol to distinguish the states $\ket{{\downarrow}{\downarrow}}$, $\ket{{\uparrow}{\uparrow}}$ and the remainder of antiparallel states.}
\end{figure}

\subsubsection{State Preparation of the Two Atoms}
\label{sec:stateprep}
Two pumping schemes are employed to prepare the atoms trapped in the cavity, one resulting in the state $\ket{{\downarrow}{\downarrow}}$ and the other one yielding an incoherent mixture of $\ket{{\uparrow}{\downarrow}}$ and $\ket{{\downarrow}{\uparrow}}$. We use a $\pi$-polarized repump laser on the $\ket{F{=}1}\leftrightarrow\ket{F'{=}2}$ transition to bring all atoms into the $\ket{F{=}2}$ ground-state manifold. For the preparation of antiparallel states, we impinge an additional right-circularly polarized pumping laser beam along the cavity axis that is resonant with the empty cavity and the $\ket{F{=}2,m_F{=}2}\leftrightarrow\ket{F'{=}3,m_F{=}3}$ transition. This pumps at least one atom into the $\ket{{\uparrow}}=\ket{F{=}2,m_F{=}2}$ state (see supplement of \cite{neuzner2016_sup}) within $170\unit{\micro{s}}$. The bare atomic transition $\ket{{\uparrow}}\leftrightarrow\ket{e}$ is red-detuned from the cavity by $80\unit{MHz}$. It is shifted into resonance via the dynamical Stark shift induced by a standing wave from a retroreflected $1.6\unit{W}$ trapping laser at a wavelength of $1064\unit{nm}$, propagating perpendicular to the cavity axis. For three-dimensional trapping, two additional, mutually orthogonal, blue-detuned lattices with a wavelength of $771\unit{nm}$ are applied. As the atoms are trapped in the nodes of these lattices, they induce no light shift. Since one atom in $\ket{{\uparrow}}$ strongly reduces the transmission of further right-circularly polarized pumping light due to the strong coupling to the cavity, the second atom remains in a different state $\ket{F{=}2,m_F{\neq}2}$ with $(86\pm4)\%$ probability. The successful pumping of at least one atom is heralded by the reduction in cavity transmission. The pumped atom is then rotated from $\ket{{\uparrow}}$ to $\ket{{\downarrow}}$ employing the Raman laser pair, thereby restoring full cavity transmission for the pumping laser.

We continue the pumping with the repump laser switched off to avoid driving of the atom in $\ket{{\downarrow}}$. A reduction in cavity transmission again heralds the successful pumping of the second atom to $\ket{{\uparrow}}$. Because we do not distinguish which of the two atoms is pumped first, the outcome of this sequence is described by the density matrix $\frac12\ket{{\uparrow}{\downarrow}}\bra{{\uparrow}{\downarrow}}+\frac12\ket{{\downarrow}{\uparrow}}\bra{{\downarrow}{\uparrow}}$. The duration of the whole preparation sequence including the pumping is $270\unit{{\micro}s}$.

To initialize the atoms to the state $\ket{{\downarrow}{\downarrow}}$, a different dipole-trap power of the $1064\unit{nm}$ trapping laser is used, such that the atoms are detuned from the empty-cavity resonance by a few MHz. With this, if one of the atoms is in $\ket{{\uparrow}}$, pumping light can still enter and the intra-cavity intensity is about half compared to an empty cavity. Therefore, pumping of the second atom continues when the first one has been prepared. After $170\unit{{\micro}s}$ of continuous pumping the state $\ket{{\uparrow}{\uparrow}}$ is prepared with $93\%$ efficiency, and a global $R^\pi_y$ rotation brings the atoms to $\ket{{\downarrow}{\downarrow}}$ (Fig.\,\figref{fig:stateprep}). To increase the overlap of the initial state of the entanglement sequence with $\ket{{\downarrow}{\downarrow}}$, we apply a heralding scheme for the state preparation. We irradiate the atoms with a resonant laser beam impinging transversally to the cavity axis, which yields fluorescence photons if at least one atom is left in $\ket{{\uparrow}}$ or any other state of the $\ket{F{=}2}$ manifold. In those cases, we discard the preparation attempt. Absence of fluorescence photons acts as a reliable herald for the preparation of $\ket{{\downarrow}{\downarrow}}$. Employing this heralding scheme, we achieve an overlap of our prepared states with $\ket{{\downarrow}{\downarrow}}$ of $99\%$.

\subsubsection{State Detection of the Two Atomic Qubits}
\label{sec:statedet}

\renewcommand{\thefigure}{S2}
\begin{figure}[b]
\centering
\positionlabel{fig:statedetection23plot}
\includegraphics[width=\columnwidth]{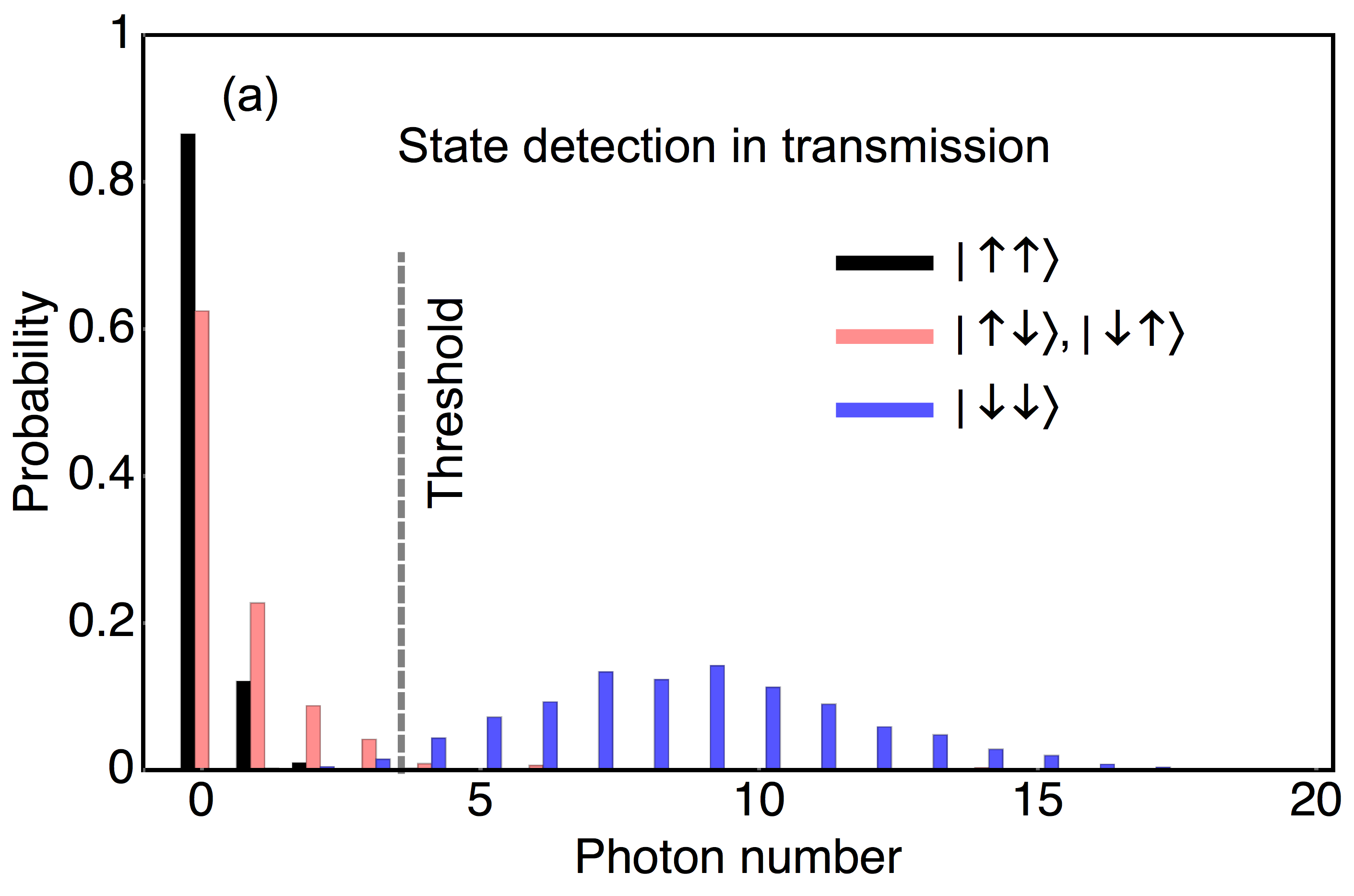}
\includegraphics[width=\columnwidth]{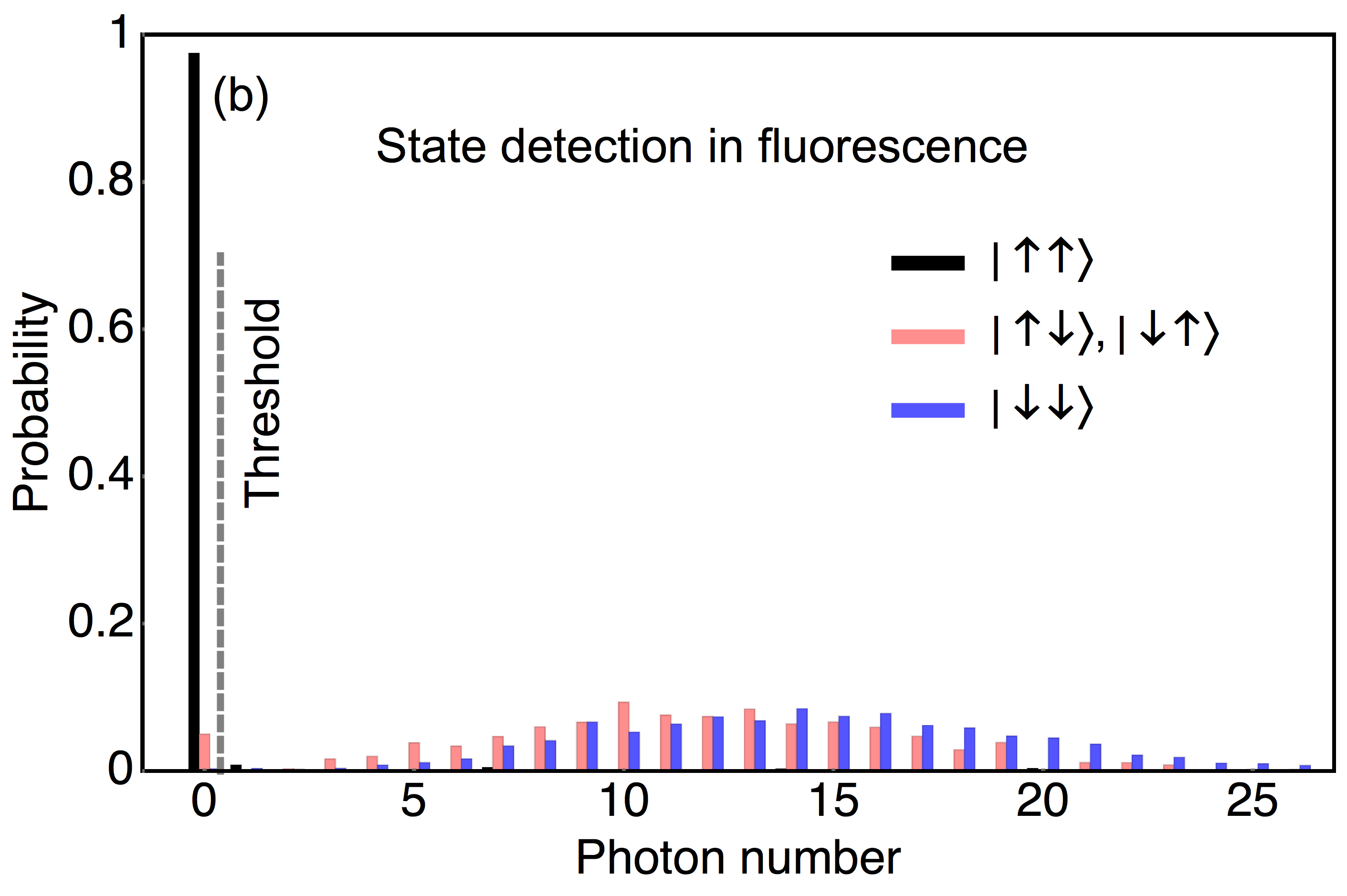}
\caption{\label{fig:statedetection23plot}
(color online). Characterization of the two state-detection techniques.
(a) Histogram of the number of detected photons for $3\unit{{\micro}s}$ of state detection in transmission. Atoms prepared in $\ket{{\downarrow}{\downarrow}}$ result in a near-Poissonian distribution (blue). Applying a criterion of more than 3 detected photons, this case can be distinguished from the distributions obtained for preparation of the atoms in $\{\ket{{\downarrow}{\uparrow}},\ket{{\uparrow}{\downarrow}}\}$ (red) and $\ket{{\uparrow}{\uparrow}}$ (black).
(b) Corresponding distributions for fluorescence state detection following a $\pi$ pulse on the two atoms. The population inversion results in a high probability of observing zero photons for atoms initially prepared in $\ket{{\uparrow}{\uparrow}}$, which we use as a discrimination criterion.}
\end{figure}

After the entangling experiment, we employ an analysis pulse with a variable phase $\phi$ and subsequently a state-detection scheme that allows us to measure the populations in the two-atom states $\ket{{\uparrow}{\uparrow}}$, $\ket{{\downarrow}{\downarrow}}$ and the sum of the populations in $\ket{{\uparrow}{\downarrow}}$ and $\ket{{\downarrow}{\uparrow}}$. The state detection starts by probing the transmission through the cavity for $3\unit{{\micro}s}$ with a right-circularly polarized laser beam resonant with the empty cavity (``Transmission'' box in Fig.\,\protect\figref{fig:stateprep}). If either of the two atoms occupies the state $\ket{{\uparrow}}$, the transmission decreases due to a normal-mode splitting of the coupled atom-cavity system. Afterwards, a Raman $\pi$ pulse (``$R_y^{\pi}$'' box in Fig.\,\protect\figref{fig:stateprep}) is applied followed by a second state-detection interval (rightmost ``Fluorescence'' box in Fig.\,\protect\figref{fig:stateprep}). This time, a laser beam resonant with the $\ket{F{=}2,m_F{=}2}\leftrightarrow\ket{F{=}3,m_F{=}3}$ transition and co-propagating with the Raman lasers irradiates the atoms for $5\unit{{\micro}s}$. This fluorescence state detection has a higher discrimination fidelity than the transmission state detection, but does not preserve the $m_F$-populations in $F=2$. It results in a near-Poissonian distribution of fluorescence photons if either of the two atoms is in $\ket{{\uparrow}}$. For both atoms in $\ket{{\uparrow}}$, the amount of the fluorescence photons strongly depends on the relative position of the atoms \cite{neuzner2016_sup, reimann2015_sup}, which may change from one experimental run to the next. Due to destructive interference of the atoms' emission, $\ket{{\uparrow}{\uparrow}}$ might even lead to fewer fluorescence photons than $\ket{{\uparrow}{\downarrow}}$ or $\ket{{\downarrow}{\uparrow}}$. Therefore, two state-detection pulses with an interleaved $\pi$ pulse are needed to discriminate between the different states with a good fidelity. Experimentally, we characterize the performance of our double-state-detection protocol in an independent measurement. For this, we infer the photon number distribution for the state detection in transmission and fluorescence for initially prepared atom-atom states $\ket{{\uparrow}{\uparrow}}$, $\ket{{\downarrow}{\downarrow}}$ and $\{\ket{{\uparrow}{\downarrow}},\ket{{\downarrow}{\uparrow}}\}$. For the state detection in transmission, the respective data is shown in Fig.\,\figref{fig:statedetection23plot}(a). We chose the criterion of more than 3 photons signalling the $\ket{{\downarrow}{\downarrow}}$ state. If 3 or less photons are observed, a distinction between $\{\ket{{\uparrow}{\downarrow}},\ket{{\downarrow}{\uparrow}}\}$ and $\ket{{\uparrow}{\uparrow}}$ is not possible. We subsequently invert the populations with the $\pi$ pulse and apply the second state detection in fluorescence. The initial state $\ket{{\uparrow}{\uparrow}}$, rotated to $\ket{{\downarrow}{\downarrow}}$ by the $R^\pi_y $ pulse, is distinct by a high probability of no detected photons as can be seen from Fig.\,\figref{fig:statedetection23plot}(b). The biggest error is a $5\%$ probability to detect no photon if the atoms were initially prepared in $\{\ket{{\uparrow}{\downarrow}},\ket{{\downarrow}{\uparrow}}\}$ (red bars). Combining the results for both state-detection methods depicted in Fig.\,\figref{fig:statedetection23plot}, we calculate the probability to detect an initially prepared state correctly. For $\ket{{\uparrow}{\uparrow}}$, this leads to less than 3 detected photons in transmission and no detected photon in fluorescence with $97.0\%$ probability. $\ket{{\downarrow}{\downarrow}}$ is indicated by more than 3 photons in transmission and at least one detection event in fluorescence in $97.4\%$ of all attempts. For $\{\ket{{\uparrow}{\downarrow}},\ket{{\downarrow}{\uparrow}}\}$, a $93.1\%$ probability to detect 3 or less photons in the first step and at least one in the second is observed. The quoted numbers are influenced by a non-perfect initial state preparation. We estimate the error of this preparation to approximately $1\%$. In our data, there is also a minor contribution of less than $0.2\%$ for that the two state detections give results incompatible with our discrimination thresholds, namely more than $3$ photons in the first state detection and no fluorescence photons in the second state detection.

\clearpage

\end{document}